\documentclass[aps,prl,showpacs,superscriptaddress,prbib, twocolumn]{revtex4}
\usepackage{graphicx}
\usepackage{float}
\usepackage{float}
\usepackage{nicefrac}
\newcommand{\Fig}[1]{Fig.~\ref{#1}}

\begin{document}

\author{Clifford W. Hicks} 
\affiliation{Scottish Universities Physics Alliance (SUPA), School of Physics and Astronomy, University of St
Andrews, St Andrews KY16 9SS, United Kingdom}
\affiliation{Max Planck Institute for Chemical Physics of Solids, N\"{o}thnitzer Str.~40, 01187 Dresden,
Germany}
\author{Alexandra S. Gibbs} 
\affiliation{Scottish Universities Physics Alliance (SUPA), School of Physics and Astronomy, University of St
Andrews, St Andrews KY16 9SS, United Kingdom}
\affiliation{School of Chemistry and EaStCHEM, University of St Andrews, North Haugh, St Andrews KY16 9ST, 
United Kingdom}
\affiliation{Max Planck Institute for Solid State Research, Heisenbergstrasse 1, 70569 Stuttgart, Germany}
\author{Lishan Zhao}
\affiliation{Scottish Universities Physics Alliance (SUPA), School of Physics and Astronomy, University of St
Andrews, St Andrews KY16 9SS, United Kingdom}
\affiliation{Max Planck Institute for Chemical Physics of Solids, N\"{o}thnitzer Str.~40, 01187 Dresden,
Germany}
\author{Pallavi Kushwaha}
\affiliation{Max Planck Institute for Chemical Physics of Solids, N\"{o}thnitzer Str.~40, 01187 Dresden,
Germany}
\author{Horst Borrmann}
\affiliation{Max Planck Institute for Chemical Physics of Solids, N\"{o}thnitzer Str.~40, 01187 Dresden,
Germany}
\author{Andrew P. Mackenzie}
\email{apm9@st-andrews.ac.uk}
\affiliation{Scottish Universities Physics Alliance (SUPA), School of Physics and Astronomy, University of St
Andrews, St Andrews KY16 9SS, United Kingdom}
\affiliation{Max Planck Institute for Chemical Physics of Solids, N\"{o}thnitzer Str.~40, 01187 Dresden,
Germany}
\author{Hiroshi Takatsu}
\affiliation{Department of Physics, Tokyo Metropolitan University, Tokyo 192-0397, Japan}
\author{Shingo Yonezawa}
\affiliation{Department of Physics, Graduate School of Science, Kyoto University, Kyoto 606-8502, Japan}
\author{Yoshiteru Maeno}
\affiliation{Department of Physics, Graduate School of Science, Kyoto University, Kyoto 606-8502, Japan}
\author{Edward A. Yelland}
\affiliation{Scottish Universities Physics Alliance (SUPA), School of Physics and Astronomy, University of St
Andrews, St Andrews KY16 9SS, United Kingdom}
\affiliation{ SUPA, School of Physics and Astronomy, and Centre for Science at Extreme Conditions, University
of Edinburgh, Mayfield Road, Edinburgh EH9 3JZ, United Kingdom }

\title{Quantum Oscillations and Magnetic Reconstruction in the Delafossite PdCrO$_2$}

\date{9 Apr 2015}

\begin{abstract}

We report quantum oscillation data on the metallic triangular antiferromagnet PdCrO$_2$. We find that, to very
high accuracy, the observed frequencies of PdCrO$_2$ can be reproduced by reconstruction of the (nonmagnetic)
PdCoO$_2$ Fermi surface into a reduced zone. The reduced zone corresponds to a magnetic cell containing six
chromium sites, giving a $\sqrt{3} \times \sqrt{3}$ in-plane reconstruction, and $\times 2$ interplane
reconstruction. The interplane ordering represents a reduction in lattice symmetry, possibly to monoclinic,
and an associated lattice distortion is expected.  In addition, we report a magnetic transition under an
applied in-plane field that is probably equivalent to the spin-flop transition reported for CuCrO$_2$, and
present data on its field-angle dependence.  We also report measurements of the resistivity of PdCrO$_2$ up to
500~K.

\end{abstract}

\maketitle

Magnetic ions coordinated on a triangular lattice often yield interesting magnetic properties. A prominent
example is the insulating CrO$_2$ sheet, stabilised in materials like LiCrO$_2$ and PdCrO$_2$. In these
compounds, the Cr formal charge is $+3$, and its configuration is $3d^3$. The crystal field at the Cr sites is
nearly octahedral, inducing a gap between the quasi-$t_{2g}$ ($d_{xy}$, $d_{xz}$, and $d_{yz}$) and
quasi-$e_g$ ($d_{3z^2-r^2}$ and $d_{x^2-y^2}$) bands. The quasi-$t_{2g}$ levels are therefore half-filled, and
the CrO$_2$ sheet is a Mott insulator: strong Hund's rule coupling aligns the spins on each Cr site, giving a
total spin on each Cr site of nearly $3/2$. The CrO$_2$ sheet is a Heisenberg system, that orders at low
temperatures into the $120^\circ$ triangular N\'{e}el phase. 

Of materials containing such a CrO$_2$ sheet, AgCrO$_2$, CuCrO$_2$, and PdCrO$_2$ have the delafossite crystal
structure, while LiCrO$_2$, NaCrO$_2$, and KCrO$_2$ have the closely-related ordered rock salt structure. All
of these systems show $120^\circ$ order, with the N\'{e}el temperature depending strongly on the Cr-Cr
spacing. The ordered rock salt structure gives substantially smaller interplane spacings than the
delafossite structure, and somewhat higher N\'{e}el temperatures.~\cite{Takatsu14}

The interlayer order is a more subtle problem than the intralayer order: in all of these systems the Cr sheets
are stacked rhombohedrally, so the molecular fields from first- and second-neighbouring layers cancel. However
the interlayer order has observable consequences: AgCrO$_2$ and CuCrO$_2$ show spin-driven ferroelectricity,
while NaCrO$_2$ and LiCrO$_2$ do not.~\cite{Seki08} AgCrO$_2$ and
CuCrO$_2$ have the delafossite structure and NaCrO$_2$ and LiCrO$_2$ the ordered rock salt structure, but the more
essential difference appears to be the interlayer order: in AgCrO$_2$~\cite{Oohara94} and
CuCrO$_2$~\cite{Kadowaki90, Frontzek12}, the interlayer ordering is ferroic, in that the vector chirality, the
rotational sense of the spin helices that comprise the  $120^\circ$ phase, is the same in all layers. In
LiCrO$_2$, in contrast, neutron scattering data suggest that the vector chirality alternates from layer to
layer~\cite{Kadowaki95}, while the interplane order of NaCrO$_2$ is not clear.~\cite{Hsieh08}

This paper focuses on PdCrO$_2$. The N\'{e}el temperature of PdCrO$_2$ is $T_N=37.5$~K, and, like LiCrO$_2$, the
vector chirality probably alternates from layer to layer,~\cite{Takatsu09, Takatsu14} so this feature is not
restricted to the ordered rock salt structure. 

Whereas the other compounds discussed above are insulators, PdCrO$_2$ is a metal, due to $4d/5s$ conduction in
the Pd sheets. The isostructural, nonmagnetic compound PdCoO$_2$ also has these Pd sheets, and its carrier
mobility was found to exceed that of copper.~\cite{Hicks12} PdCrO$_2$ therefore is an interesting system,
comprised of highly-conducting sheets interleaved with Mott insulating spacer layers. It would be interesting
to determine whether the metallic Pd conduction has any effect on the magnetic order.

What is certainly true is that the Pd conduction can be used as a probe of the magnetic order.
Quantum oscillation~\cite{Ok13} and angle-resolved photoemission spectroscopy (ARPES)~\cite{Sobota13, Noh14}
studies have shown that the PdCrO$_2$ Fermi surfaces result from reconstruction of the PdCoO$_2$ Fermi surface
into the magnetic zone. Observation of an unconventional anomalous Hall effect in PdCrO$_2$ indicates further
that the Cr spins are not co-planar.~\cite{Takatsu10UAHE}

In this paper we present measurements of quantum oscillations in PdCrO$_2$, and confirm the overall frequencies
reported in Ref.~\cite{Ok13}. We add greater resolution and a careful comparison with the nonmagnetic
PdCoO$_2$ Fermi surface: we show that reconstruction of the PdCoO$_2$ Fermi surface reproduces most of the
observed oscillation frequencies to very high accuracy; the PdCrO$_2$ frequencies can be analyzed in detail
without recourse to density functional theory calculations. We show that the magnetic coupling is
$k_z$-dependent: it is much weaker at $k_z=\pm \pi$ than at $k_z=0$. Also, the dominant magnetic scattering
vectors are those corresponding to a 6-Cr magnetic unit cell giving a $\times 2$ interplane reconstruction in
addition to the $\sqrt{3} \times \sqrt{3}$ in-plane reconstruction. The interplane order represents a
reduction in lattice symmetry, from $R\bar{3}m$ (rhombohedral) to, in the highest-symmetry case, $C2/c$
(monoclinic).

We also report a magnetic transition under applied field, at a similar field to a spin flop transition
reported in CuCrO$_2$, that was observed in the course of measurement of the oscillations. Finally, we present
measurements of the resistivity of PdCrO$_2$ to high temperatures.

\section{Methods}

Single crystals of PdCrO$_2$ were grown by a NaCl flux method, using PdCrO$_2$ powder synthesized via a
solid-state reaction.~\cite{Takatsu10crystalGrowth} We measured magnetic oscillations in two samples of
PdCrO$_2$, by torque magnetometry, using the same piezoresistive AFM cantilevers as in our previous study on
PdCoO$_2$.~\cite{Hicks12, Cantilever} The cantilevers were mounted on rotatable platforms with integrated
field angle sensors. 

\begin{figure}[ptb] \includegraphics[width=3.25in]{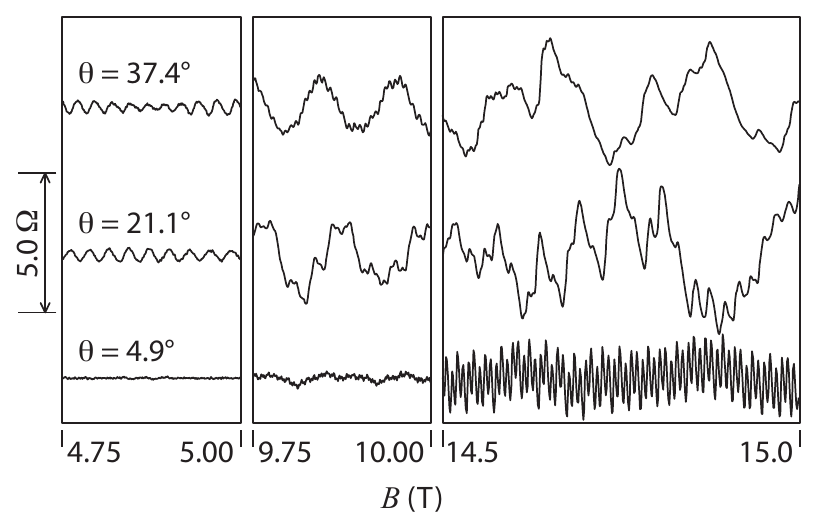}
\caption{\label{exampleOscillations} Magnetic oscillations in sample \#1 at three field
angles.  The $y$-scale is the resistance of the piezoresistive sense element on the cantilever.} \end{figure}

Sample \#1 was roughly $230 \times 300 \times 11$~$\mu$m, and sample \#2 $160 \times 100 \times 15$~$\mu$m.
For sample \#1 the field was rotated about a $\langle 1000 \rangle$ axis (using hexagonal indexing), and for
sample \#2 a $\langle 1 \bar{1} 00 \rangle$ axis; these axes are illustrated at the bottom right of
\Fig{frequencies}.
Some raw data for sample \#1 are shown in
\Fig{exampleOscillations}. Low-frequency oscillations ($\sim$800~T for field angle $\theta \rightarrow 0$)
dominate the data. At 0.7~K, these oscillations were discernible at fields as low as 2~T. At high fields
(above $\sim$10~T), the oscillation amplitude was very large, leading to strong torque interaction: the
oscillations had a triangular form, and strong sum and difference frequencies appeared in the Fourier
transforms. While torque interaction complicates the analysis somewhat, it may be difficult to avoid if the
samples are to be large enough for higher-frequency oscillations to be resolved.

The oscillation amplitudes were history-dependent. Cooling the sample through $T_N$ in a 15~T field resulted
in oscillation amplitudes for the $\alpha$, $\beta$, and $\gamma$ orbits roughly four times as large as with
zero-field cooling. These large amplitudes persisted as long as the field was maintained. Releasing the field
and heating the sample to $\sim$1~K caused the amplitudes to decrease, and re-applying fields up to 15~T did
not recover the large amplitudes. The system therefore appears to be nonergodic. A possible origin of glassy
behaviour is domain-boundary Cr spins: the relatively large magnetic unit cell of PdCrO$_2$ means that a
complex domain structure is likely.

For measurement of the temperature dependence of the oscillation amplitudes, we cooled the sample in a field
and kept the field above 8~T at all times, to maintain large oscillation amplitudes. For measurement of
the angle dependence of the frequencies, torque interaction was a greater concern, and the sample
was cooled through $T_N$ in zero field.~\footnote{While measuring the angle dependence of the oscillations in
sample \#1, we performed a ``de-Gauss'' routine at the start of each run, in which the field was ramped to
-2~T, then +1~T, then -0.5~T, and so on. The aim was to reduce the probability of the system evolving
gradually into a more ordered state, with larger oscillation amplitudes, though it is not clear whether the
procedure had much effect.}

\section{Results: the oscillations}

\begin{figure}[ptb]
\includegraphics[width=3.25in]{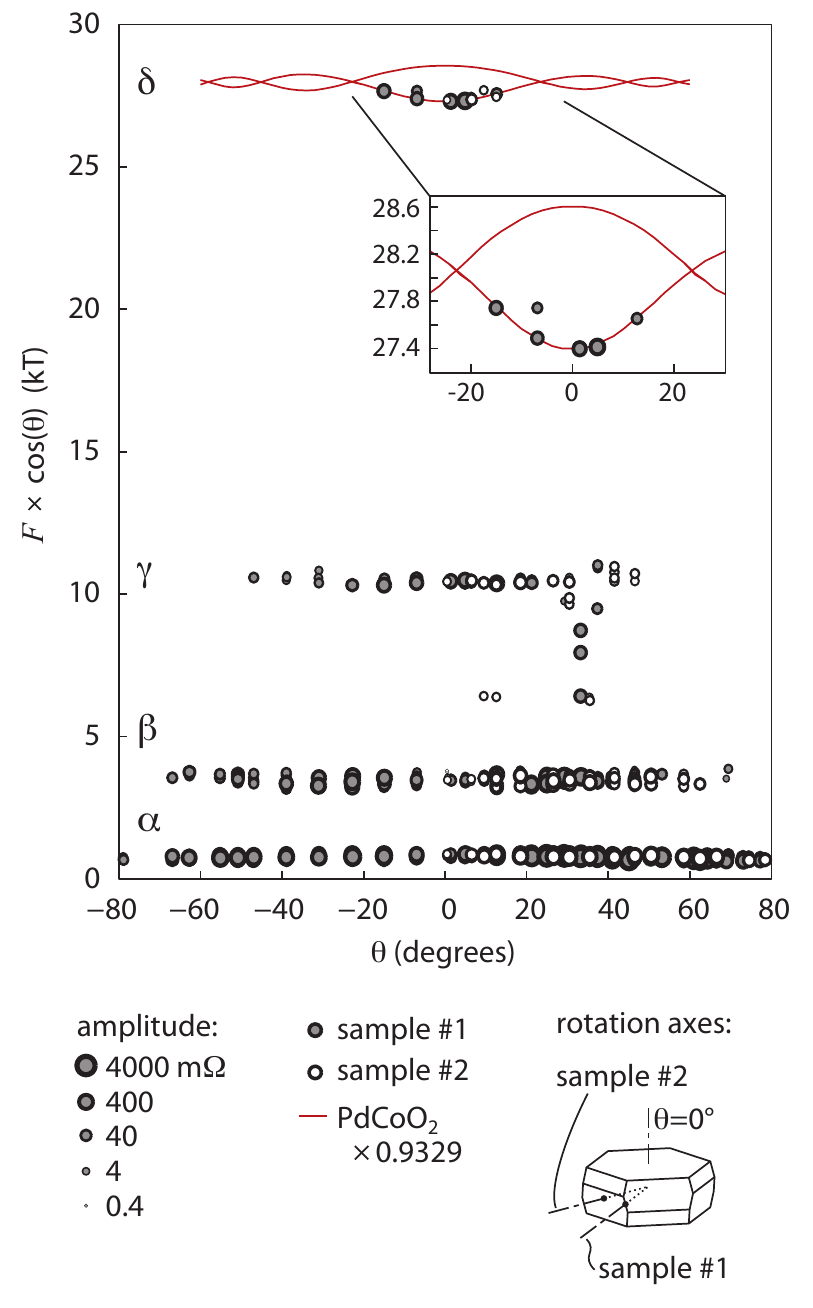}
\caption{\label{frequencies} The observed oscillation frequencies of PdCrO$_2$, multiplied by $\cos(\theta)$,
against field angle $\theta$. The field rotation axes are shown at bottom right, referenced to the nonmagnetic
zone. The frequencies are from Fourier transformation of the data over the range 4.2 to 15~T. Frequencies that
are clearly sum or difference frequencies are not shown. The indicated amplitudes of the oscillations are the
peak-to-peak amplitudes of the oscillations in resistance of the sense element on the
cantilever. The PdCoO$_2$ frequencies are taken from the model in Ref.~\cite{Hicks12}, and have been scaled by
93.3\%.}
\end{figure}

The oscillation frequencies as a function of field angle are shown in \Fig{frequencies}.  Four sets of
intrinsic peaks can be identified, located, at $\theta=0$, at around 0.8, 3.5, 10.5, and 27.4~kT. Following
Ref.~\cite{Ok13}, we label these $\alpha$, $\beta$, $\gamma$, and $\delta$. All the frequencies scale broadly
as $1/\cos(\theta)$, indicating that the Fermi surfaces are highly two-dimensional.

\begin{figure}[ptb]
\includegraphics[width=3.25in]{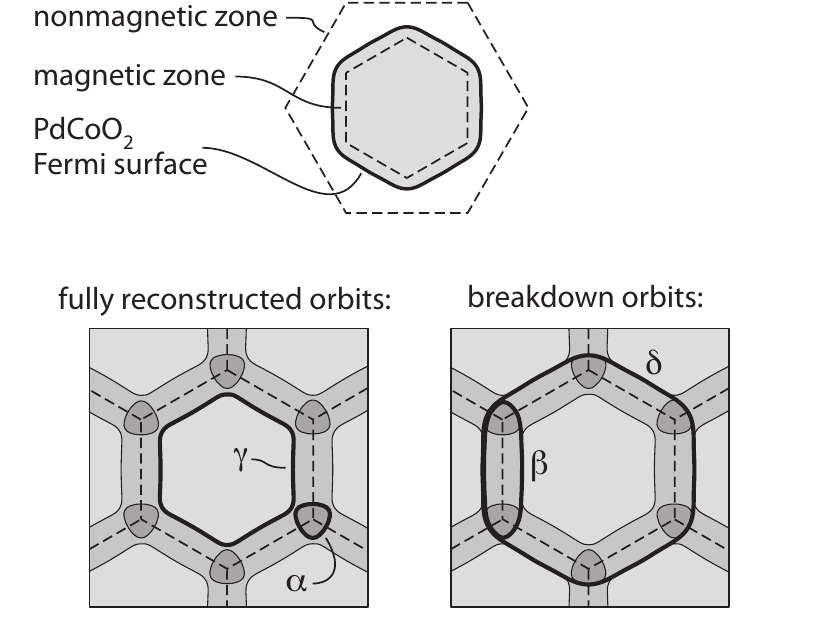}
\caption{\label{reconstruction} A two-dimensional model of the reconstruction. Top: the PdCoO$_2$ Fermi
surface in the 2D nonmagnetic zone, and magnetic zone arising from a $\sqrt{3} \times \sqrt{3}$
reconstruction. Bottom: reconstruction into the magnetic zone. The fully reconstructed orbits, $\alpha$ and
$\gamma$, are illustrated in the left-hand panel, and the breakdown orbits, $\beta$ and $\delta$, in the
right-hand panel.}
\end{figure}

It has been established that the PdCrO$_2$ orbits result from a $\sqrt{3} \times \sqrt{3}$ reconstruction of
the PdCoO$_2$ Fermi surface, due to the 120$^\circ$ N\'{e}el order.~\cite{Ok13, Noh14} This reconstruction is
illustrated in \Fig{reconstruction}. The $\alpha$ and $\gamma$ orbits are fully reconstructed orbits, while
$\beta$ and $\delta$ result from magnetic breakdown, {\it i.e.} the orbits cross gaps in $k$-space.

The cyclotron masses were determined for sample \#1 by Lifshitz-Kosevich fits to the temperature dependence of
the amplitudes. We analysed the data between 7.5 and 11.5~T, avoiding higher fields where torque interaction
was very strong, and verified that within this range there was no systematic variation of the masses with
field. 

The masses obtained for the $\alpha$, $\beta$, $\gamma$, and $\delta$ orbits are $0.33 \pm 0.01$, $0.84 \pm
0.01$, $1.37 \pm 0.02$, and $(1.55 \pm 0.04)m_e$, respectively, the fits are shown in the Appendix.  These are
in good overall agreement with ARPES measurements. A Fermi velocity of 4.2~eV-\AA{} was measured, by ARPES, at the
corners of the nonmagnetic Fermi surface, and 4.9~eV-\AA{} at the faces.~\cite{Noh14} If an isotropic
Fermi velocity of 4.6~eV-\AA{} is taken along the perimeters of the $\alpha$, $\beta$, $\gamma$, and $\delta$
orbits, masses of 0.34, 0.63, 1.00, and $1.53m_e$ are obtained, respectively: the cyclotron masses from the
Lifshitz-Kosevich fits are in very close agreement with the ARPES estimates for $\alpha$ and $\delta$, and
$\sim$35\% heavier for $\beta$ and $\gamma$.~\footnote{For calculation of the $\alpha$ and $\gamma$ expected
masses, we use the $\alpha_3$ and $\gamma_1$ orbits.}

Also shown in \Fig{frequencies} are the PdCoO$_2$ frequencies from the parametrised model in
Ref.~\cite{Hicks12}, for comparison with PdCrO$_2$. To make the comparison, the PdCoO$_2$ frequencies need to
be scaled by the square of the ratio of in-plane lattice constants. We found that the best match is
obtained with a scaling of 93.3\%, which is very close to the expected scaling, (2.830~\AA{}/2.923~\AA{})$^2$
= 93.7\%.~\cite{Takatsu14, Takatsu07} 

From the comparison it is apparent that the lower branch of the PdCoO$_2$ frequencies, arising from the neck
orbit ({\it i.e.} the $k_z=\pm \pi$ orbit), is visible in the PdCrO$_2$ data, while the upper branch, from the belly
orbit ($k_z=0$), is not.
This feature is part of a pattern that extends to the other frequencies: the observed breakdown frequencies
derive from reconstruction of the nonmagnetic neck orbit, while the observed
fully-reconstructed frequencies derive mainly from the belly orbit. The pattern is illustrated in
Table~I, which shows the observed PdCrO$_2$ frequencies (for $\theta \rightarrow 0$) and the expected
frequencies based on reconstruction of the PdCoO$_2$ neck and belly orbits: the neck reconstruction yields the
observed $\beta$ and $\delta$ frequencies, and the belly reconstruction $\alpha_3$ and $\gamma$. (The $\alpha$
frequencies comprise three sub-bands, $\alpha_1$, $\alpha_2$, and $\alpha_3$; we will discuss this in more
detail below.) 

What this pattern means is that the belly orbit sees the magnetic order more strongly than the neck orbit.
This is not surprising: the Cr sites are midway between the Pd sheets (which dominate conduction), and the
$k_z=\pm \pi$ Bloch states have zero weight at the planes of the Cr nuclei.
\begin{table}[ptb]
\begin{tabular*}{\columnwidth}{@{\extracolsep{\fill}}llllll}
\hline\hline
 			      & $\alpha_3$ & $\beta$   & $\gamma$   & $\delta$ \\
\hline
\rule{0pt}{1.5ex}observed frequencies (kT): 	        & 0.87 	& 3.45      & 10.52      & 27.40 \\
\hline
\rule{0pt}{1.5ex}calc'd $k_z=0$ freq's (kT):   	   & \bf{0.91}  & 3.92 	    & \bf{10.50} & 28.60 \\
calc'd $k_z=\pi$ freq's (kT):  & 0.69 	& \bf{3.36} & 11.27      & \bf{27.40} \\
\hline
\end{tabular*}
\caption{Observed frequencies at $\theta=0$, and expected frequencies from two-dimensional $\sqrt{3} \times
\sqrt{3}$ reconstruction of the PdCoO$_2$ $k_z=0$ and $k_z=\pm\pi$ orbits. Boldface indicates a better match
to observations. In the calculations, the PdCoO$_2$ frequencies are scaled by 93.3\%, and avoided crossings
at band crossings are neglected apart from their effect on Fermi surface topology.}
\end{table}

Overall, then, a two-dimensional $\sqrt{3} \times \sqrt{3}$ reconstruction of the PdCoO$_2$ frequencies,
where the reconstruction is weak at $k_z=\pm \pi$, gives a good description of the PdCrO$_2$ frequencies at
$\theta=0$. We now test the reconstruction at other field angles by performing the full three-dimensional
reconstruction.

\begin{figure}[ptb]
\includegraphics[width=3.25in]{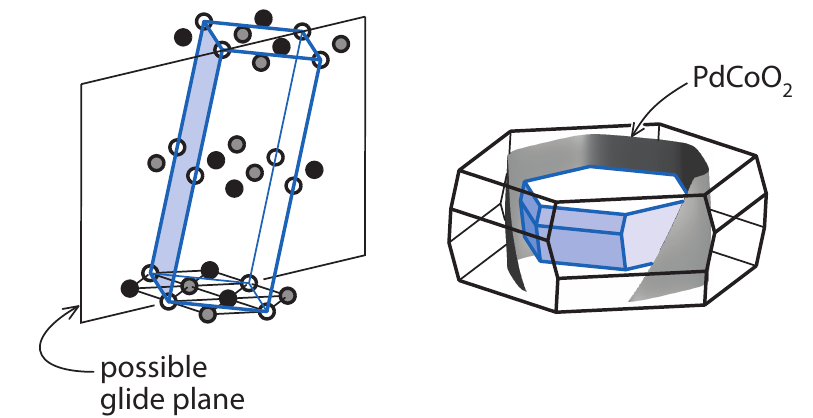}
\caption{\label{3Dreconstruction} 
Left: a possible magnetic cell for PdCrO$_2$, that contains 6 Cr sites. The Cr sites are colored black, gray,
and white to indicate the three spin directions of each layer. Right: the nonmagnetic first Brillouin zone
(black), together with the PdCoO$_2$ Fermi surface and the 6-site zone, into which the nonmagnetic Fermi
surface is reconstructed.}
\end{figure}

We first need a three-dimensional magnetic cell. It must contain at least six Cr sites: three per layer to
capture the $120^\circ$ N\'{e}el order, and two layers to capture the alternating vector chirality. The cell
we test is shown in \Fig{3Dreconstruction}.~\footnote{In Refs.~\cite{Kadowaki95} (LiCrO$_2$)
and~\cite{Takatsu14} (PdCrO$_2$), single-crystal neutron scattering data are analyzed using an 18-site
magnetic cell. This is the smallest that can preserve $R\bar{3}m$ symmetry: to preserve $R\bar{3}m$ symmetry
the interplane ordering vector must take on all three possibilities in an ABCABC order, so six layers are
required to capture both this and the alternating vector chirality. The resulting most-likely magnetic
structure in Ref.~\cite{Takatsu14} can be reduced to a six-site magnetic cell.} The Cr sites are stacked
rhombohedrally, so at $T_N$ there are three equivalent choices for the interplane ordering vector. In choosing
one of them the lattice symmetry is reduced. The space group of the nonmagnetic lattice is $R\bar{3}m$. The
highest-symmetry possible magnetic lattice has space group $C2/c$ (base-centered monoclinic): reflection about
the glide plane indicated in \Fig{3Dreconstruction} reverses the vector chirality within each layer, and
translation along the interplane ordering vector restores the original lattice.~\footnote{The unit cell shown
in \Fig{3Dreconstruction} is metrically triclinic --- none of the angles are $90^\circ$ --- but if the glide
symmetry is preserved then a 12-site monoclinic supercell can be constructed.} Magnetostructural coupling
should lead to a lattice distortion associated with the reduction in symmetry, and it is possible that
more subtle features in the magnetic order reduce the symmetry further.

The reconstruction is performed as described in textbooks: sections of the nonmagnetic Fermi surface are
translated by combinations of reciprocal lattice vectors of the reduced zone until all portions of the
original surface are represented within the reduced zone. For our nonmagnetic surface, we take the parametrised
model for the PdCoO$_2$ Fermi surface that was determined in Ref.~\cite{Hicks12}. We do not include finite
avoided crossings in our calculation, although the topology of the reconstructed Fermi surfaces is determined
by the orientations of the avoided crossings that would occur in the real system. Further details of the
calculation are given in the appendix. Our results are shown in \Fig{matchingTheFrequencies}, together with
the observed frequencies. 

There are two
$\gamma$ bands. The lower band, $\gamma_1$, derives from reconstruction of the belly orbit, and $\gamma_2$ the
neck orbit. $\gamma_2$ is not observed in the data because the magnetic coupling is weak for the neck orbit.
\begin{figure}[ptb]
\includegraphics[width=3.25in]{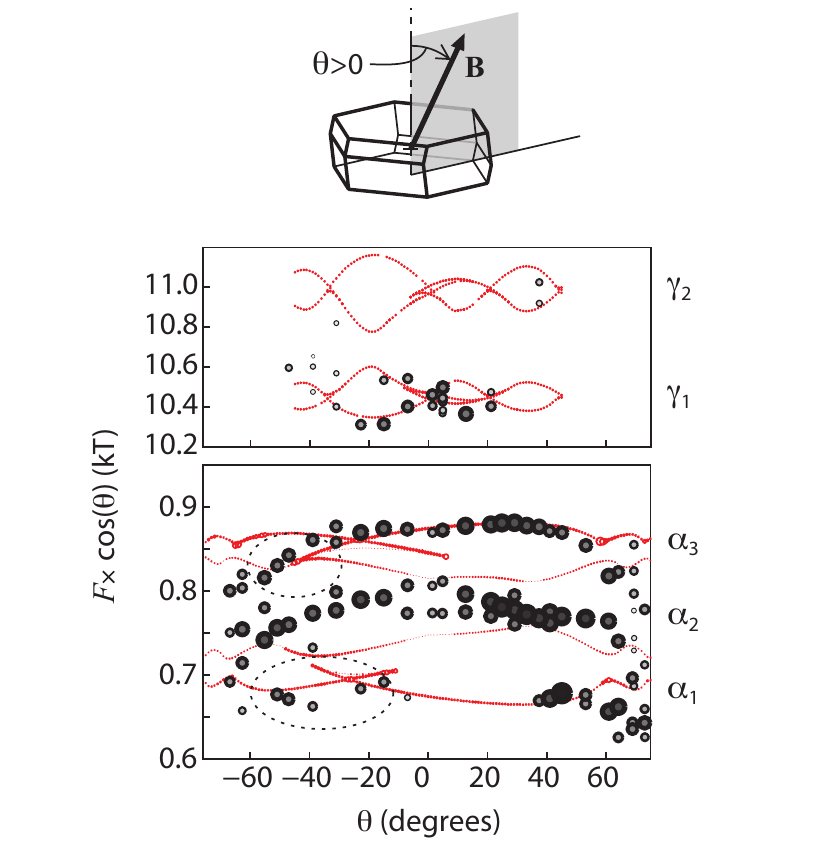}
\caption{\label{matchingTheFrequencies} Red circles: calculated frequencies based on reconstruction of the
PdCoO$_2$ Fermi surface into the magnetic zone shown in \Fig{3Dreconstruction}. The upper and lower
panels show the $\gamma$ and $\alpha$ frequencies, respectively. In the lower panel, the diameter of the
symbols is proportional to the logarithm of the expected oscillation amplitude, based solely on the curvature
of the Fermi surfaces. Gray circles: the frequencies observed in sample \#1, with the diameter of the symbols
proportional to the logarithm of the amplitude.}
\end{figure}

Three $\alpha$ bands appear in the data.~\footnote{Ok {\it et al} also identify three $\alpha$
bands.~\cite{Ok13} However the frequencies they label $\alpha_1$ and $\alpha_2$ are both part of $\alpha_2$ in
our labelling scheme, and Ok {\it et al} do not resolve the frequency we label $\alpha_1$.} The lower and
upper bands, $\alpha_1$ and $\alpha_3$, are reproduced well by the calculation. $\alpha_3$ is from the
nonmagnetic belly orbit and $\alpha_1$ the neck orbit, so $\alpha_1$ has a much lower amplitude in the data.
The middle band, $\alpha_2$, is not reproduced in the calculations. It could be a breakdown orbit. There are
prominent breakdown orbits in the data ($\beta$ and $\delta$), and the $\times 2$ interplane reconstruction
could lead to breakdown orbits that mix segments from the original neck and belly orbits.

For this calculation we supposed that the system chose the interplane ordering vector that aligned the plane of
(possible) glide symmetry with the field rotation plane. The other two possibilities would lead to the glide 
and rotation planes being separated by $120^\circ$. We also calculated this possibility, with the result 
shown in the Appendix. The match to the data is reasonable but not as good, so it appears that either the
glide and rotation planes were aligned by chance, or the fields applied during the measurement re-oriented
the magnetic reconstruction.

We also show in the Appendix results for a strictly two-dimensional reconstruction, that would preserve the
$R\bar{3}m$ symmetry of the nonmagnetic lattice. The results do not match the data well, and we conclude that
the dominant magnetic scattering vectors are those of the magnetic cell shown in \Fig{3Dreconstruction}.

\section{Spin flop transition}

\begin{figure}[ptb]
\includegraphics[width=3.25in]{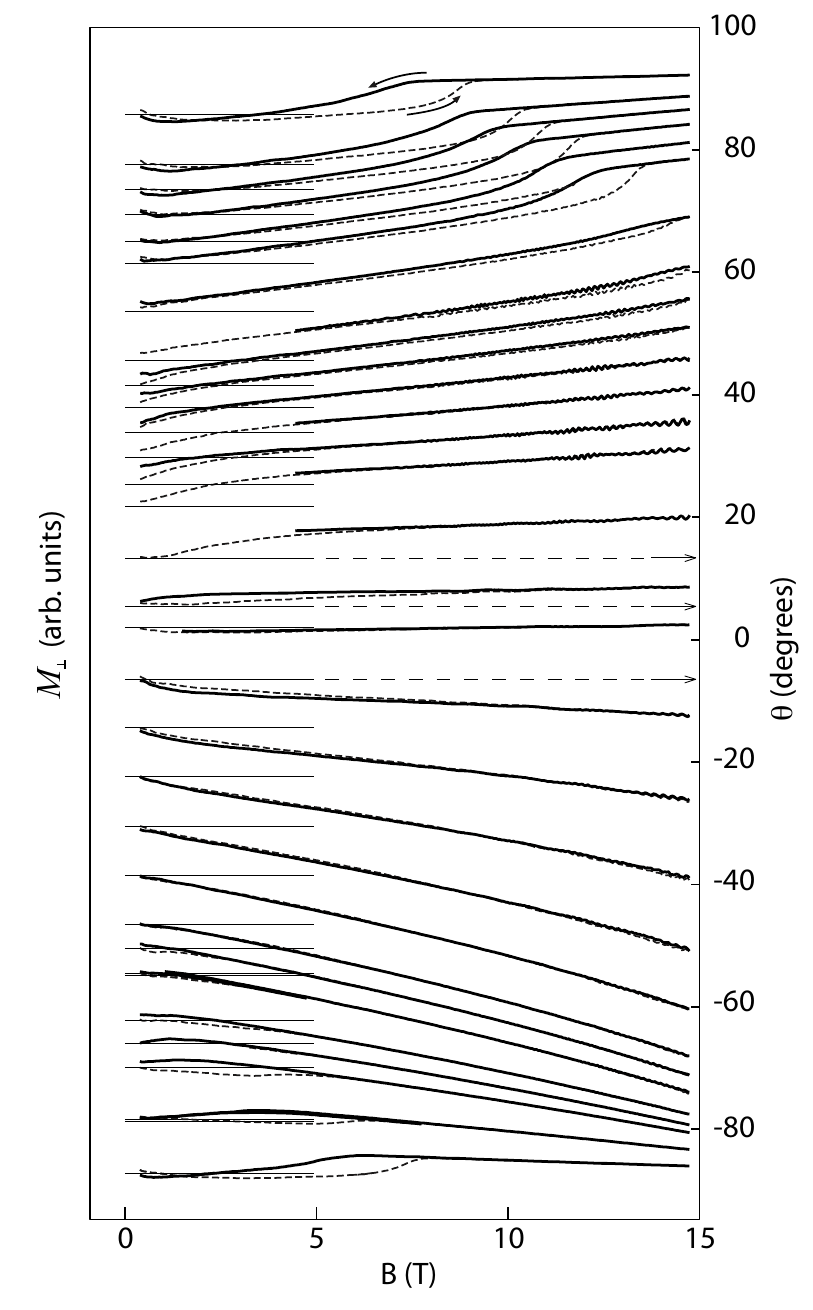}
\caption{\label{Mperp} Perpendicular magnetisation against applied field at 0.7~K, for
sample \#1, where the field was rotated in a $\langle 0001 \rangle$-$\langle 1\bar{1}00 \rangle$ plane. Each
curve is offset by the field angle $\theta$; the offsets are indicated by the thin horizontal lines at the
left and the angle scale is on the right. The solid lines show downward field sweeps, the dashed lines
upward.}
\end{figure}

In addition to quantum oscillations, a large-scale feature appeared during the torque magnetometry
measurements on sample \#1: a first-order magnetic transition.  It occurs at a field of around 6.5~T for
$\theta=90^\circ$ ({\it i.e.} the field applied in the plane). On the high-field side of the transition the
sample has a much larger $c$-axis magnetic moment than on the low-field side. \Fig{Mperp} shows the set of
torque curves obtained in this study, divided by the applied field to yield $M_{\perp}$, the magnetisation
perpendicular to the applied field, and smoothed so as to exclude the oscillations and show the broad-scale
features. The transition and its hysteresis are readily apparent, at $\theta \sim 90^\circ$. It appeared for
sample \#1 but not \#2, so it is sensitive to the direction of the applied field.

Over a small angle range, quantum oscillations were visible both below and above the transition; the Fourier
transforms are shown in the Appendix. At each angle within this range, the oscillations have a lower amplitude
on the high-field side of the transition than on the low-field side, and the dominant peak shifts to a
slightly higher frequency.  That the transition affects the oscillations in a consistent manner shows that it
is a bulk property.

This transition is at a similar field to a spin-flop transition reported for CuCrO$_2$,~\cite{Kimura09_PRL}
5.3~T for temperatures well below $T_N$, and probably has the same origin. In CuCrO$_2$, it has been found by
single-crystal neutron diffraction that the spins lie in a $\langle 0001 \rangle - \langle 1\bar{1}00 \rangle$
plane.~\cite{Soda10, Poienar10, Frontzek12} The transition occurs when the field is applied in a $\langle 1
\bar{1}00 \rangle$ direction, but there is no transition for fields applied in a $\langle 1000 \rangle$
direction. It has been shown by symmetry arguments~\cite{Kimura09_PRL} and by direct observation~\cite{Soda10}
that the transition is a spin flop, where the spin plane rotates by $90^\circ$.

For PdCrO$_2$, although observation of an unconventional anomalous Hall effect shows that the spins are not
co-planar, neutron scattering data show that the spins lie approximately in a $ \langle 0001 \rangle-\langle
1\bar{1}00 \rangle$ plane, similar to CuCrO$_2$. Furthermore, and also as in CuCrO$_2$, the transition occurs
when the field is applied in a $\langle 1\bar{1}00 \rangle$ direction (sample \#1), but not in a $\langle 1000
\rangle$ direction (\#2).  Therefore the transition is probably the same spin flop.

\section{Resistivity}

PdCoO$_2$ has been used as a nonmagnetic analogue of PdCrO$_2$ in order to extract the magnetic contribution
to the specific heat and electrical resistivity.~\cite{Takatsu10JPhys, Takatsu09} We have shown here that it
is a very good comparison: the PdCrO$_2$ Fermi surfaces are, to high precision, a reconstruction of the
PdCoO$_2$ Fermi surface, and the cyclotron masses match closely: $(1.55\pm0.04)m_e$ for the $\delta$ orbit of
PdCrO$_2$, against $1.45\pm0.05m_e$ for the (equivalent) neck orbit of PdCoO$_2$.~\cite{Hicks12} 
The in-plane resistivity of PdCoO$_2$ is substantially non-linear between $\sim$100 and 500~K, a feature
attributed to prominent optical phonons.~\cite{Takatsu07} The resistivity of PtCoO$_2$ also shows this
feature.~\cite{Kushwaha14} We include in this paper data on the resistivity of PdCrO$_2$ up to 500~K, in order
to extend the comparison reported in Ref.~\cite{Takatsu10JPhys} to higher temperatures and to see whether the
same feature appears in PdCrO$_2$.

\begin{figure}
\includegraphics[width=3.25in]{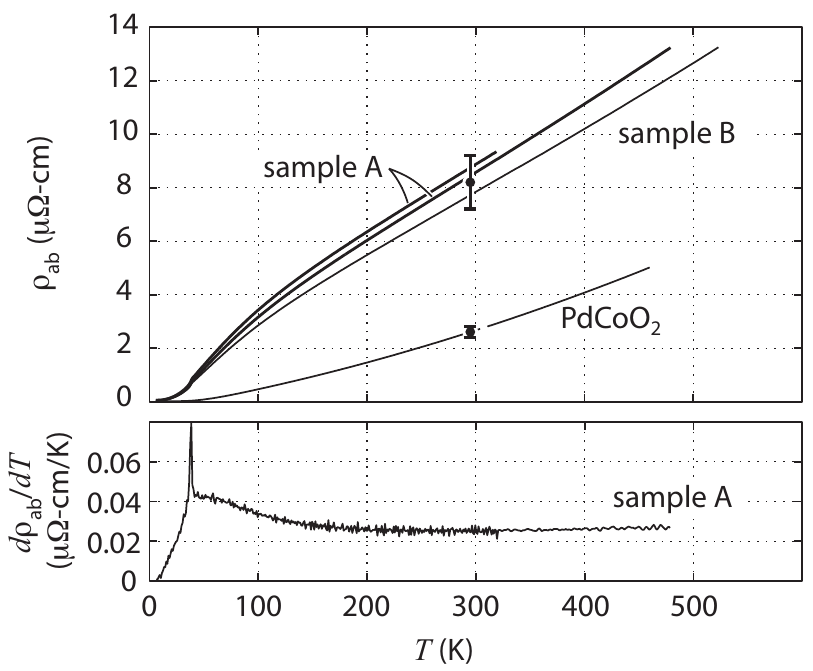}
\caption{\label{resistivity} The resistivity of PdCrO$_2$ against temperature, for two samples. Sample A had
two pairs of voltage contacts. The point at 295~K, with error bars, indicates the room-temperature
resistivity: $8.2\pm1.0$~$\mu\Omega$-cm. Data on a PdCoO$_2$ sample are also shown, scaled to the
room-temperature resistivity determined in Ref.~\cite{Hicks12}: $2.6 \pm 0.2$~$\mu\Omega$-cm. Bottom panel:
$d\rho/dT$ for sample A.}
\end{figure}

All samples for resistivity measurement were cut with a wire saw into bars of nearly constant width and thickness, and with
length-to-width ratios of $\sim$10, to reduce geometrical uncertainties in conversion of resistance to
resistivity. Our data are plotted in \Fig{resistivity}. We measured one PdCoO$_2$ and two PdCrO$_2$ samples,
labelled A and B; sample A had two pairs of voltage contacts, so in total three PdCrO$_2$ curves were recorded.
Averaging the three measurements, the room temperature (295~K) resistivity of PdCrO$_2$ was found to be
$8.2\pm1.0$~$\mu\Omega$-cm, where the uncertainty is from uncertainty in the sample dimensions. This is in
good agreement with that reported in Ref.~\cite{Takatsu10JPhys} (9.4~$\mu\Omega$-cm). 

The form of the resistivity of PdCrO$_2$ is well-established. There is a sharp cusp at $T_N$, and above $T_N$ the
magnetic component of the resistivity, $\rho_m$, remains well below its saturation value due to short-range
correlation. The increase and eventual saturation of $\rho_m$ as the temperature increases leads to a convex
temperature dependence of the resistivity.

The in-plane resistivity of PdCrO$_2$, in contrast to PdCoO$_2$ and PtCoO$_2$, is essentially linear from
$\sim$200 to at least 500~K. It may be that the temperature dependence of $\rho_m$ obscures an optical phonon
contribution. $\rho_m$ is expected to saturate when the Cr spins become completely uncorrelated. If $\rho_m$
is estimated by subtracting the resistivity of PdCoO$_2$ from that of PdCrO$_2$, then our data indicate that
$\rho_m$ continues to increase at temperatures well above room temperature, {\it i.e.} $\rho_m$ continues to
evolve to temperatures an order of magnitude greater than $T_N$. The Weiss temperature of PdCrO$_2$ is
$\sim$-500~K,~\cite{Takatsu09} so correlations between the Cr spins are expected to persist to temperatures up
to $\sim$500~K, and a high saturation temperature of $\rho_m$ may be expected.

\section{Discussion and Conclusion}

The quantum oscillation frequencies indicate that the dominant magnetic scattering vectors are those of a
six-site magnetic cell with lower symmetry than the nonmagnetic lattice. An associated lattice distortion is
expected to onset at $T_N$. It has been looked for by both neutron and X-ray diffraction, and not
found,~\cite{Takatsu14} so any distortion must be small.

The distortion, if it occurs, should resemble that observed in CuCrO$_2$;~\cite{Kimura09} there is
also evidence for a lattice distortion in AgCrO$_2$.~\cite{Lopes11} In CuCrO$_2$, the vector chirality is the
same in each layer and the magnetic cell contains three sites.~\cite{Seki08, Kadowaki90, Frontzek12} The magnetic
transition is split into two,~\cite{Aktas13} which is expected for easy-axis-type $120^\circ$
antiferromagnetism: spins first order along the easy axis, then, at a lower temperature, along an in-plane
axis. Ultrasound velocity measurements show that the lattice distortion starts at the upper transition, with
no apparent anomaly at the lower transition.~\cite{Aktas13}.

The transition of PdCrO$_2$ may also be split, although if so the splitting is very small and not generally
observed in experiment.~\cite{Takatsu14} Either way, the magnetic order implies two separate reductions in
symmetry from $R\bar{3}m$: the choice of interplane ordering vector at the (possible) upper transition, and
the choice of spin plane at the lower transition (where, again, in PdCrO$_2$ the spins are only approximately
co-planar). The two symmetry reductions are likely to couple. The ultrasound data on CuCrO$_2$ suggest that
in that system it is the interplane ordering vector that more strongly drives the distortion. 

The angle dependence of the spin-flop transition in PdCrO$_2$ is interesting: the transition field appears to
vary smoothly from $\sim$15~T on one side of the plane (at $\theta \sim +55^\circ$) to nearly zero on the
other (at $\theta \sim -55^\circ$). At $B=0$, however, the field angle is a meaningless parameter, so if there
is in fact an endpoint near $B=0$ then the high- and low-field states are adiabatically connected to each other.
Whether this is true and how it might occur for a $90^\circ$ spin flop requires further investigation.

In summary, PdCrO$_2$ is an interesting system comprised of alternating highly-conductive sheets and
Mott-insulating spacer layers. It provides a good model for study of the interaction between metallic
conduction and antiferromagnetic order: the conduction is well-characterized, arising from a single, open
Fermi surface, and much is known about the magnetic order. Also, there are related compounds that permit
precise comparisons, such as the nonmagnetic PdCoO$_2$, and other CrO$_2$-based delafossite and ordered rock salt
materials. Several avenues of inquiry remain open.

We acknowledge useful discussions with John W. Allen, Christopher A. Hooley, Peter Thalmeier, and Burkhardt
Schmidt, and practical assistance from Nabhanila Nandi. We acknowledge funding from the UK EPSRC, the
MEXT KAKENHI (No. 21340100), the Royal Society, the Wolfson Foundation, and the Max Planck Society.

\section{Appendix}

In the Appendix we first present some supplementary data, and then give further details on the calculation of
the reconstruction.

\begin{figure}[b]
\includegraphics[width=3.25in]{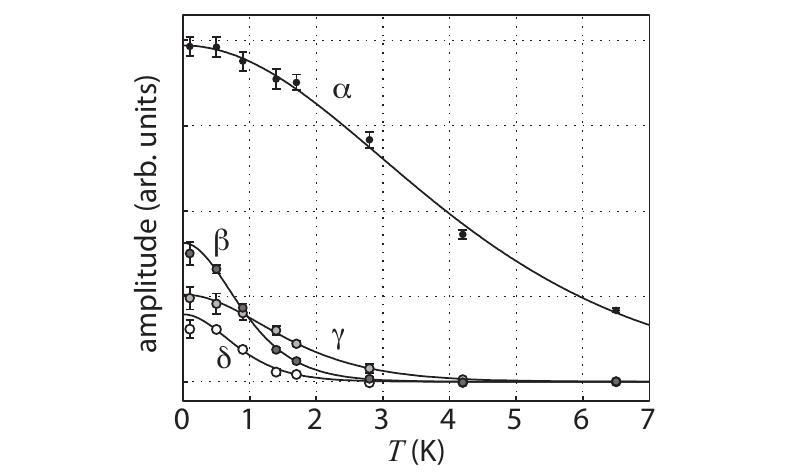}
\caption{\label{LKfits} Lifshitz-Kosevich fits to the amplitudes determined over the field range 8.5--9.5~T.}
\end{figure}

\Fig{LKfits} shows the Lifshitz-Kosevich fits to the oscillation amplitudes determined over the field range
8.5--9.5~T. Four such fits were done, over 1~T field ranges starting at 7.5, 8.5, 9.5, and 10.5~T. The masses
reported in the main text are the averages of the masses obtained from these fits.

\begin{figure}
\includegraphics[width=3.25in]{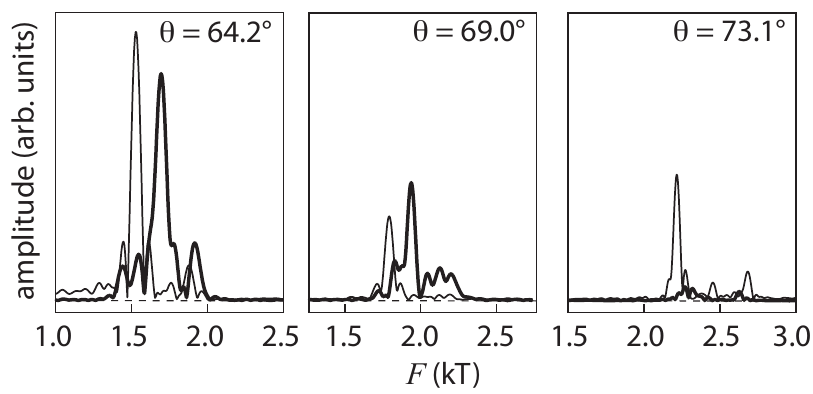}
\caption{\label{aboveAndBelow}Fourier transforms of the oscillations below (thin lines) and above (thick
lines) the magnetic transition, which at the angles in this figure occurs at $B\sim10$~T.}
\end{figure}

\Fig{aboveAndBelow} shows the Fourier transforms of the oscillations above and below the putative spin-flop
transition, for the narrow angle range over which oscillations were observed on both sides of the transition.
On the high-field side of the transition, the oscillations have lower amplitudes and the main peak a higher
frequency. (The reduction in amplitude may not be immediately apparent in the figure, but normally
quantum oscillation amplitudes grow rapidly as the field is increased, and the reduction in amplitude is very
clear in the raw data.)

We now give further details on the reconstruction. The first step is to determine the misalignment, if any,
between the sample and the field angle sensor mounted on the rotator platform. This misalignment is
determined, for sample \#1, by comparison of the observed $\delta$ frequencies and the scaled PdCoO$_2$
frequencies, shown in \Fig{frequencies}. The best match is obtained when a $0.3 \pm 0.2^\circ$
offset is added to the measured field angle. This is our misalignment, and correction for it is incorporated
into all plots of sample \#1 frequencies presented in this paper.

We take the nonmagnetic Fermi surface to be the parametrised PdCoO$_2$ Fermi surface that was determined in
Ref.~\cite{Hicks12}. The parameters specifying the Fermi surface are the radius of a circular, cylindrical
base Fermi surface, $k_{00}$, and then the amplitudes of the corrugations on this base. $k_{60}$, for example,
is the amplitude of the hexagonal distortion. $k_{01}$ sets the difference between the radii of the neck and belly
orbits. A thorough description of this system of parametrization is given in Ref.~\cite{Bergemann03}.

We allow three parameters to be adjusted, within small ranges that are consistent with experiment, to match
the overall levels of the observed PdCrO$_2$ frequencies. We emphasize that this adjustment only tunes the
overall levels, and does not substantially alter any of the substructure.  These parameters are: (1) The
overall size of the Fermi surface ($k_{00}$), (2) The magnitude of the hexagonal distortion ($k_{60}$), and
(3) the in-plane lattice constant, $a$. The $k_z$-dependent Fermi surface corrugations ($k_{01}$, $k_{02}$,
and $k_{31}$) are left fixed at the amplitudes found for PdCoO$_2$. When $k_{60}$ is adjusted, the
higher-order amplitudes $k_{12,0}$ and $k_{18,0}$ are adjusted by hand in response, to smooth the faces and
sharpen the corners of the Fermi surface so as to match the Fermi surfaces observed by ARPES.~\cite{Sobota13,
Noh14} 

The best match is found with $k_{00}=0.9219$~\AA{}$^{-1}$, $k_{60}=0.036$~\AA{}$^{-1}$, and $a=2.927$~\AA{}.
Further refinement (for example by allowing a low-temperature lattice distortion, or modification of the
$k_z$-dependent corrugations, or finite avoided crossings) might yield slightly different results--- the
obtained lattice parameter, for example, is not precisely in line with expectations, as it has been measured
to be 2.923~\AA{} at room temperature and some thermal contraction is expected.~\cite{Takatsu14} But these
fitting parameters are within bounds permitted by experiment and are adequate to model the angle dependence.

\begin{figure}
\includegraphics[width=3.25in]{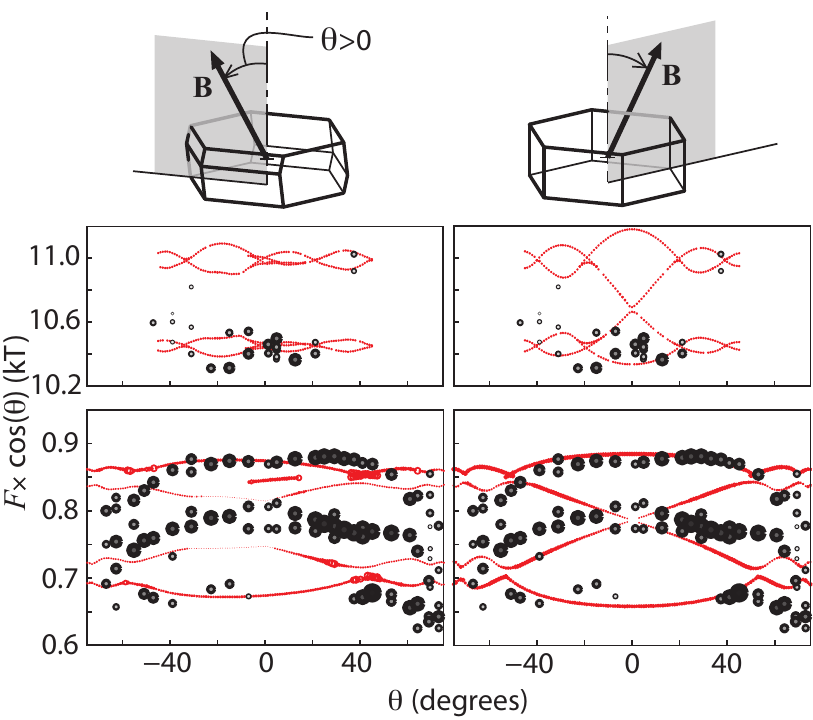}
\caption{\label{matchingTheFrequencies_appendix}  Oscillation frequency against field angle for two further
possibilities for the magnetic reconstruction, discussed in the text.}
\end{figure}

In addition to the reconstruction shown in the main text, we tested two further possibilities. The first is
the same reconstruction but with the other possible orientation relative to the field rotation plane, that the
glide plane is rotated from the field rotation plane by $120^\circ$. The second tests a hypothetical
magnetic zone with vertical side walls, that would preserve the $R \bar{3} m$ symmetry of the nonmagnetic
lattice. The results are shown in \Fig{matchingTheFrequencies_appendix}.

The rotated reconstruction, shown in the left-hand panels in \Fig{matchingTheFrequencies_appendix}, gives a
reasonable match to the data, but not as good as with the glide and rotation planes aligned: it yields the
wrong sign on the slope of $\alpha_3$ against $\theta$, and too low an amplitude on the substructure of the $\gamma_1$
frequencies. The $R\bar{3}m$ reconstruction gives a worse match: although it reproduces the $\alpha_1$ and
$\alpha_3$ frequencies reasonably well, it does not reproduce the structure on $\gamma_1$ at all. Therefore,
we conclude that the magnetic reconstruction illustrated in Figs.~\ref{3Dreconstruction} and
\ref{matchingTheFrequencies} is the correct one.


\begin{thebibliography}{99}
\bibitem{Cantilever}The cantilevers are model PRC120, from Seiko Precision Instruments. The sense element is
piezoresistive, and the datasheet lists the response, at room temperature, as 16.5~$\Omega$/$\mu$m. The spring
constant is $\approx$40~N/m, and the length 120~$\mu$m; these parameters can be combined to yield a response
to torque of $\approx 3 \cdot 10^9$~$\Omega$/N-m.
\end{thebibliography}
\end{document}